\documentclass[10pt,sigconf]{acmart}

\usepackage{graphicx}
\usepackage{url}
\usepackage{xcolor}

\acmDOI{10.1145/XXXX}

\acmISBN{978-1-4503-XXXX-X}

\acmConference[OpenSuCo'17]{Workshop on Open Source Supercomputing}{November 2017}{Denver, Colorado USA} 
\acmYear{2017}
\copyrightyear{2017}

\acmArticle{4}
\acmPrice{15.00}

\begin{document}


\title[Evaluating a Lookup Accelerator with Emulation and Simulation]{Combining Emulation and Simulation to Evaluate a Near Memory Key/Value Lookup Accelerator}

\author{Joshua Landgraf}
\affiliation{%
  \institution{Lawrence Livermore National Laboratory, Livermore, CA}
}
\email{landgraf6@llnl.gov}

\author{Scott Lloyd}
\affiliation{%
  \institution{Lawrence Livermore National Laboratory, Livermore, CA}
}
\email{lloyd23@llnl.gov}

\author{Maya Gokhale}
\affiliation{%
  \institution{Lawrence Livermore National Laboratory, Livermore, CA}
}
\email{gokhale2@llnl.gov}


\maketitle

\begin{abstract}
  Processing large numbers of key/value lookups is an integral part of
  modern server databases and other ``Big Data" applications. Prior
  work has shown that hash table based key/value lookups can benefit
  significantly from using a dedicated hardware lookup accelerator placed near
  memory. However, previous evaluations of this design on the Logic in
  Memory Emulator (LiME) were limited by
  the capabilities of the hardware on which it was emulated,
  which only supports a single CPU core and a single
  near-memory lookup engine. We extend the emulation results by
  incorporating simulation to evaluate this design
  in additional scenarios. By incorporating an HMC simulation model,
  we design optimizations that better mitigate the effects of the
  HMC closed page policy and that better utilize the HMC's parallelism, 
  improving predicted performance by an order of magnitude.
  Additionally, we use simulation to evaluate the scaling
  performance of multiple near-memory lookup accelerators. Our work
  employs an open source emulator LiME, open source simulatation
  infrastructure SST, and the open source HMC-Sim simulator.

\end{abstract}

%
%

\small


%
%

%
%



\normalsize

\vspace{-.1in}
\section{Introduction}
\label{sec:intro}
The emergence of heterogeneous architectures incorporating accelerators, combined with new memory technologies, presents a unique opportunity to explore novel, high performance node and memory architectures. Modeling, simulation, and emulation play key roles in assessing the impact of architectural variations on applications and workloads. Rapid exploration of alternative designs enables quantitative identification of both bottlenecks and opportunities relative to selected metrics.

In this work we combine hardware emulation and software simulation to study a near memory acceleration module accessing memory with varying latency profiles. Emulation gives nanosecond-scale precision in designing components of a single near-memory key/value store lookup accelerator with a simplified memory model. Software simulation is then used to study the effects of other memory models and the performance of multiple lookup accelerators. These additional scenarios help us identify optimizations to the hardware lookup engine that simulation indicates could give an order of magnitude performance improvement in lookups/sec on a Hybrid Memory Cube. Additionally, we leverage the speed advantage of the emulator in exploring several design parameters with the flexibility a software simulation that can scale the design to multiple units. This allows the simplified software simulation to run at a reasonable speed and additionally allows for the integration of external modules not available in the emulation framework.

\section{Approach}
Our simulation builds on the prior work of Lloyd and Gokhale \cite{Lloyd2017} to design and test a near-memory key/value lookup accelerator. We use SST \cite{sst1} and HMC-Sim \cite{leidel2016hmc} to develop a simulation of the hardware lookup accelerator, replicate the accelerator, and integrate it with a Hybrid Memory Cube (HMC) simulation component.

\subsection{Key/Value Store}
The key/value store is organized as an open-address hash table \cite{introalgs}. In open addressing, the entire table is allocated at initialization, and collisions are resolved by inserting colliding entries close to the hashed table slot. The number of entries to search to resolve a lookup request is called the {\em probe sequence length} and this length depends on the percent of occupied slots in the hash table, the {\em load factor}. The lookup engine is agnostic to the insertion scheme as long as the colliding entry is inserted near the hashed table slot (i.e. it is not re-hashed to a completely different location). To resolve a lookup request, it starts at the hashed location and searches a probe sequence length number of entries. The lookup engine is also agnostic to characteristics of the query stream, specifically, how often a specific key appears in the query stream. The query repeat frequency does however affect CPU performance, as frequently repeated keys are more likely to be found in the CPU cache and not require memory requests. Evaluation of the lookup accelerator in the emulator includes both uniform random key query sequence as well as Zipfian distribution ($f_k \propto k^{-\alpha}$) with $\alpha=.99$. 

\subsection{Lookup Accelerator}
\label{acc}
\begin{figure}
\begin{centering}
\includegraphics[width=\columnwidth]{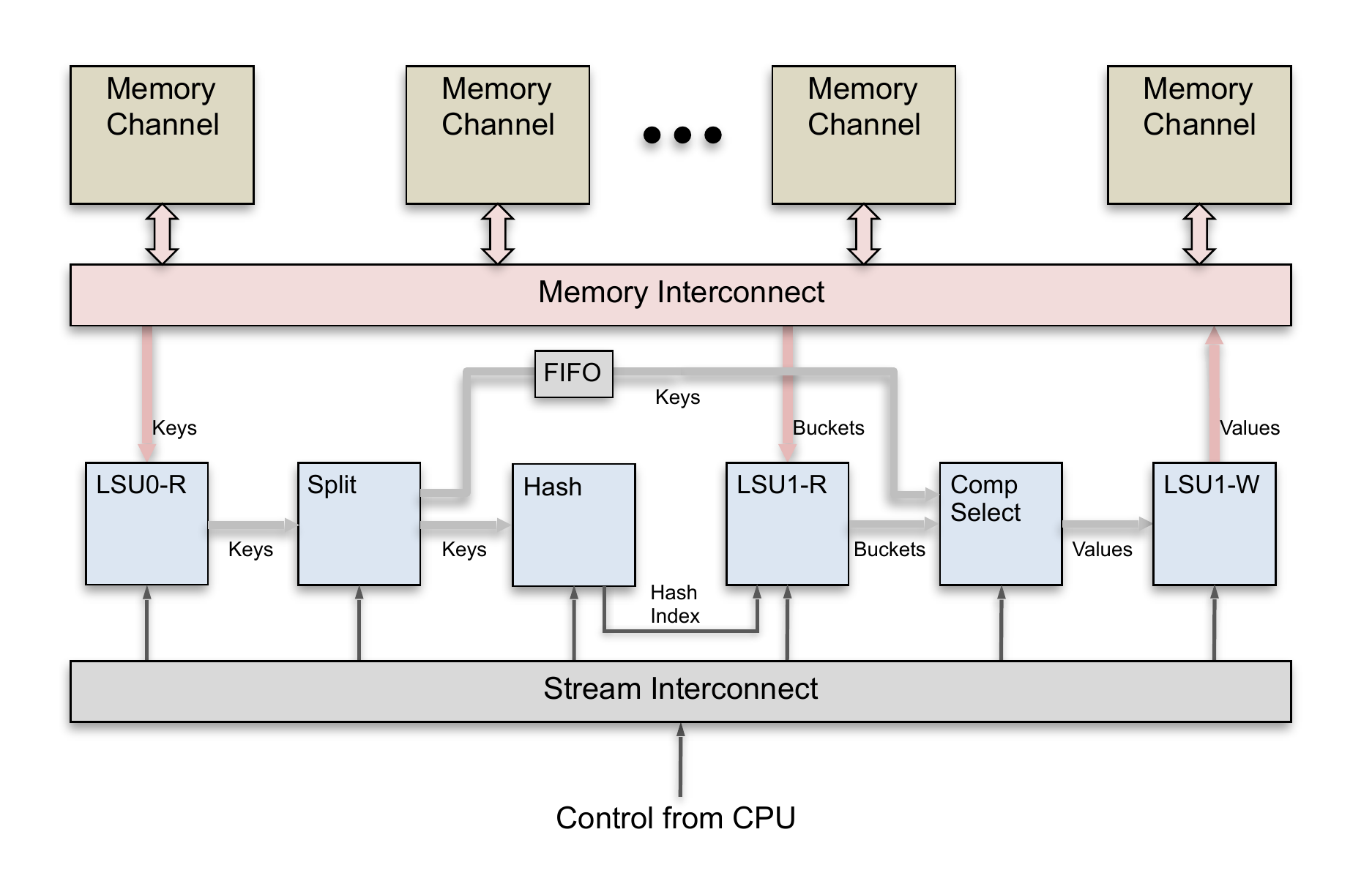}
\par\end{centering}
\caption{\label{fig:pipeline} Lookup pipeline}
\end{figure}

The original lookup accelerator by Lloyd et al. was designed for processing batched hash table lookups at high throughput on an HMC. The layout of the accelerator is detailed in Figure~\ref{fig:pipeline} and was implemented on the LiME open source emulator \cite{lime} for performance analysis. It consists of a collection of modules in a pipeline, including a hash unit, multiple load/store units (DMA blocks) \cite{lloyd2015}, a module to pass the data stream to two destinations (splitter), and a compare/select unit to compare the target key with the key stream and recover the value. The LSU can either read or write to sequential, strided, or random locations, as directed by a control stream. In the diagram, narrow lines indicate control streams, and thicker gray lines represent data streams.

In operation, the CPU loads a batch of keys into scratchpad memory and signals the lookup engine to begin. The first load/store unit (LSU), LSU0-R, reads in keys from memory and outputs them to the splitter, which then passes them to the FIFO and hash unit. The hash unit processes the keys into indices and passes them to another LSU, LSU1-R, which reads a probe sequence from the hash table at the provided index. LSU1-R outputs the probe sequence to the compare/select unit (CSU), which compares the keys in the probes to the next key from the FIFO. If there is a match, the corresponding value is output to the final LSU, LSU1-W, which writes the value to an SRAM scratchpad (not shown). If no match is found in the entire probe sequence, the CSU outputs a ``key not found" value instead. Once the batch of lookups is complete, the accelerator signals the CPU that it is done, and the CPU reads the values from the accelerator's scratchpad memory.
In the simulation, this hardware design is coded in software as an SST component.

\subsubsection{SST}
The simulation is implemented in the Structural Simulation Toolkit (SST) \cite{sst1,sst2,sst3}. SST is a powerful tool for implementing simulations for a number of reasons. It combines synchronous and event-driven simulation models, allowing for components to communicate across clock domains. SST also comes with a large range of features: the core library includes a set of tools for building components while the elements library contains a suite of prebuilt components. These range from statistics counters and variable-latency links to cache and memory simulators. SST can also interface with external libraries and includes support for libraries like DRAMSim2 \cite{dramsim2}, METIS \cite{metis}, and HMC-Sim \cite{leidel2014hmc,leidel2016hmc}. Finally, SST is designed to be scalable through the use of thread-safe data structures, MPI communication libraries, and support for checkpointing.

SST is an evolving open source project, with contributions from both the original developers as well as the research community. To date, by SST's evaluation metrics, most of the components in the elements library are considered to be in sandbox phase. SST includes some high level CPU and cache models and provides hooks to integrate external contributions of additional component simulations. For example, the well known DRAM simulator DRAMSim2 can be included in an SST simulation. Similarly, gem5 \cite{gem5-2011,gem5-2015} models can be incorporated into an SST simulation. gem5 includes detailed CPU and GPU models that can be adapted for architecture exploration. SST continues to be modified and improved, which has resulted in significant changes to part of its core infrastructure. Newer versions of SST might introduce modifications to the component interface, requiring porting of existing simulations to new APIs.

\subsubsection{HMC-Sim}
\label{HMC-Sim}
As one experiment in our lookup accelerator simulator, we simulate a near memory hardware module in which the memory is a Hybrid Memory Cube. For this purpose, we incorporate HMC-Sim into our SST simulation. The HMC-Sim memory model follows the Hybrid Memory Cube specification \cite{hmc-spec1, hmc-spec2}. It models a simple DRAM with additional infrastructure to support routing, queuing, and reordering of packets from links to quads and from quads to vaults.

HMC-Sim has several different versions that support corresponding versions of the HMC specification: HMC-Sim v1 is based on HMC spec 1.0--1.1, and v2 and v3 are based on HMC spec 2.0--2.1. Since the constants used throughout HMC-Sim are based on these specifications, it is difficult to use newer versions of HMC-Sim to simulate HMCs based on the 1.0--1.1 spec (like Micron's HMCs). For this reason, unless the source code is modified, a particular simulation can only modify a small number of supported configuration options based on the version of HMC-Sim in use. In the version of HMC-Sim used in this work, some configuration options are not implemented in the actual simulation. For instance, the link speed macro sets a register that is otherwise unused in the simulation. HMC-Sim also simplifies simulation of the movement of data within the HMC. Instead of transferring data in the granularity of flow units (FLITs) as described in the HMC spec, HMC-Sim moves whole packets at a time. This simplification extends to simple memory accesses. Entire read and write operations are assumed to complete in a fixed latency, regardless of the read or write size. Furthermore, HMC-Sim assumes that entire queues can be processed in a single clock cycle.

\subsection{Contributions}
The major contribution of our work is to demonstrate how the design of a detailed hardware emulation may be incorporated into a software simulator, translating the speed advantage of the emulator into an abstracted software simulation that can run at a reasonable speed and incorporate components not available in the emulator. Using LiME, SST, and HMC-Sim, we have 
\begin{itemize}
\item designed a detailed model of Lloyd and Gokhale's lookup accelerator and a simple supporting CPU in SST
\item developed a memory model in SST similar to the one in LiME and used it to verify the simulation
\item incorporated HMC-Sim into our simulation to test performance on a more complex HMC memory model
\item quantified the impact of accelerator design optimizations not present in the original hardware design
\item estimated through simulation how the optimized accelerator should scale on future HMC designs.
\end{itemize}

\section{Simulation Design}
\label{sec:design}

Our simulation includes three main components: a memory manager, a lookup accelerator, and a CPU. In order to make sure our models were accurate, we used our knowledge of the original emulator design to create a high-level layout and filled in lower-level implementation details by analyzing patterns in memory traces gathered from the emulator. To facilitate testing of our design, we parameterized many of these lower-level details.

\subsection{Memory Manager}
The memory manager allows the simulation to access memory with a consistent interface. The memory requests are then processed by either the internal emulator memory model or external HMC-Sim model. Memory requests can be generated internally in the CPU component or can be read from a memory trace file produced by LiME. 

\subsubsection{Emulator Memory Model}
The LiME infrastructure enables precise control over memory latency at .25\,ns granularity through the use of programmable delay units that enforce the desired latency. Maximum bandwidth is determined by the clock frequency and width of the emulated interconnect to memory. The emulator has sufficient bandwidth to allow a single accelerator to run without restriction. The simulator implements a fixed-latency memory accessed over a fixed-bandwidth link. The fixed latency is directly modeled in the SST infrastructure through a delay. To simulate a limited-bandwidth link, we serialized data going over the link and limited the rate of transfer to the bandwidth of the link. This not only simulates the time it takes to transfer the data but also simulates the throttling of memory requests whose rate would exceed maximum bandwidth.

\subsubsection{HMC-Sim Memory Model}
When using the HMC-Sim memory model, our memory manager forwards accesses to HMC-Sim through SST's memHierarchy interface. We modified the original interface to preserve access sizes since memHierarchy otherwise assumes all accesses are of the same size, which is not true for the accelerator. It is worth noting that this does not currently affect performance due to the way HMC-Sim tracks data movement as described in Section \ref{HMC-Sim}.

\subsection{Lookup Accelerator}
\label{acc-design}
At a high level, our accelerator model is close to the original design (Section \ref{acc}). Since run times are mainly influenced by memory performance, we faithfully model the details of the Load Store Units, including the memory request rate and queue depth based on those observed in the emulator's timestamped memory traces.

We also added support for three optimizations to the original design. In the original version, the read request for each key in the query is sent individually to the vault containing the group of query keys. With a closed page policy, this results in a delay when the stream of key read requests is issued. As an optimization, we send read requests for multiple keys in a single 128B packet (16 keys in a request for 8B keys). Thus, instead of reading keys from the vault individually, we can obtain a group of keys in one request using the largest packet size the memory supports. This significantly reduces overheads and eliminates the high number of bank delays caused by the base scheme. We also allowed for doubling the width of the accelerator data path so that it would only take 1 cycle to read in a hash table entry instead of two. Finally, we also doubled the number of outstanding memory requests that the memory system could support since this seemed to be limiting throughput in the original design as well.

\subsection{CPU}
To support the accelerator, we include a CPU model in our simulation. The CPU's main role is to perform cache flushes and invalidations, configure the accelerator, and read values back from the scratchpad. We created a simple CPU component that mainly simulates the timing effects of these operations. In these experiments, there is one CPU per accelerator. The CPU models support dynamic load balancing so that they will continue to provide their accelerators with work until none is left.

\subsection{Configurability}
The simulation provides several configuration options. Both CPU and accelerator clock frequencies can be set. While the accelerator's layout is fixed, most components can be configured to take different numbers of cycles. For instance, the hash unit's and compare/select unit's pipeline delays are configurable, along with various delays associated with the LSUs. The accelerator's scratchpad memory has a configurable fixed-time latency as well. The simulation can also handle changes to the size of keys, values, hash table entries, and HMC packets, which are helpful in exploring different hash table layouts and newer HMC devices. The simulation supports many of the options from the emulator as well, including lookup batch size, probe sequence length, and hash table size. Finally, the accelerator model can be configured to simulate different hash table access patterns. The default behavior is random uniform accesses, but emulator traces captured through real runs of meaningful data can also be simulated.

\section{Experiments and Results}
\label{sec:exps}

We performed four major experiments with our simulation to validate its accuracy and test how the accelerator would perform under a variety of previously untested scenarios. For all of these experiments, we varied the load factor of the hash table, which impacts the probe sequence length by causing more collisions as the table becomes more full. We also configured the accelerator with a memory trace from the emulator to further match the access patterns of the genome data used and its distribution (0.99 zipf).

We report two different performance timings in our results. ``Lookup" measures the time from when the CPU starts flushing keys from its caches to when it invalidates the caches for the values in the accelerator's scratchpad. This is the time reported in Lloyd and Gokhale's original results. ``Full lookup" also includes the time taken for the CPU to read back the values from the scratchpad. Neither method of timing includes the time taken by the CPU to write the keys to memory as this varies with different workloads.

\subsection{Verification}
\begin{figure}
\begin{centering}
\includegraphics[width=\columnwidth]{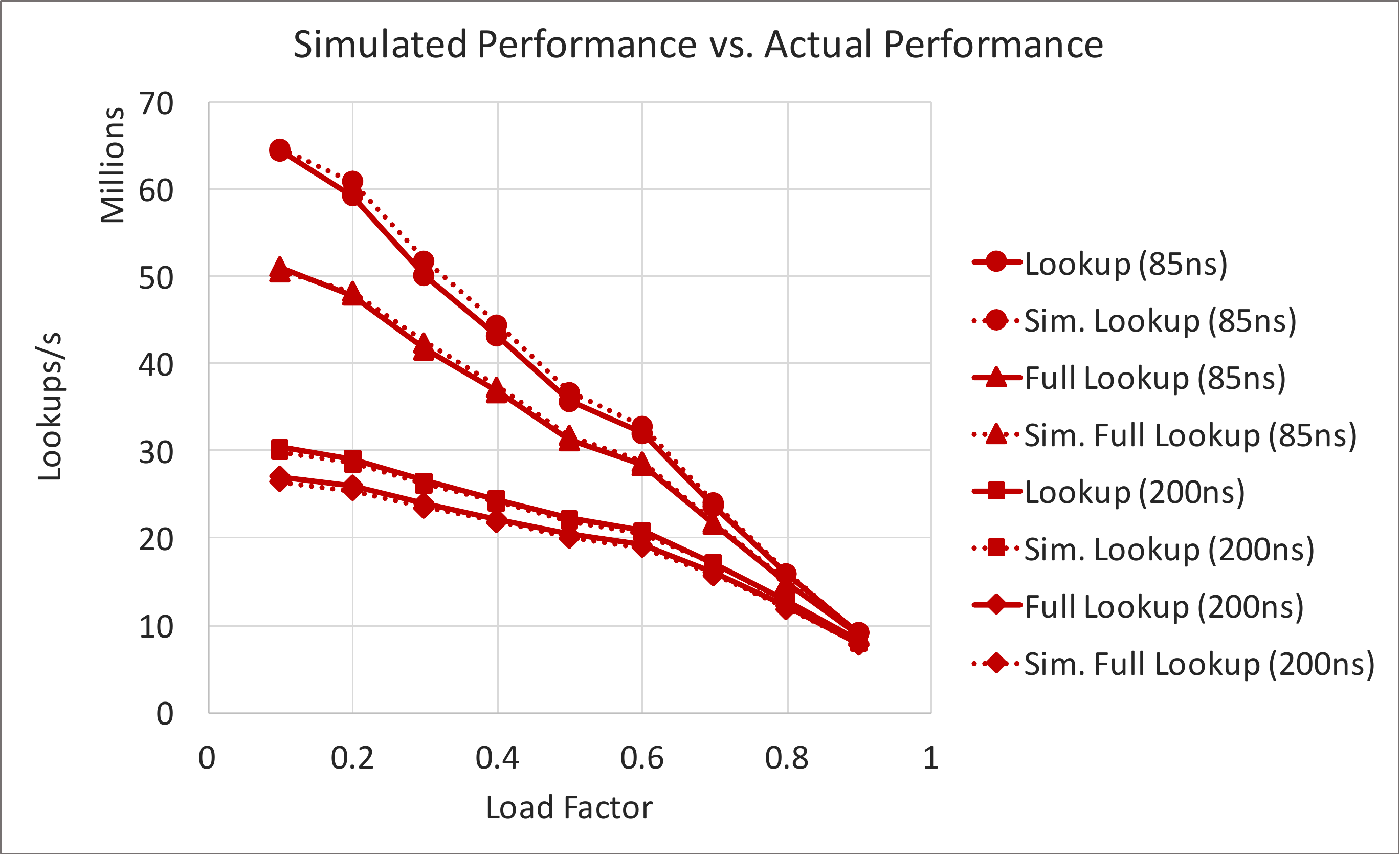}
\par\end{centering}
\caption{\label{fig:verification} Verification results}
\end{figure}

In order to verify the accuracy of our simulation models, we configured the simulator to be close to the emulator's configuration using our emulator memory model designed specifically for this purpose. We configured the memory to use the same read latencies from the original results: 85\,ns for HMC/HBM and 200\,ns for Storage Class Memory, and we limited the bandwidth of the link between the accelerator and memory to 10\,GB/s. The results are shown in Figure~\ref{fig:verification}.

With these configuration parameters, our simulation matches the emulator results to within 4\%, with an average error of 1.7\%. This gives us confidence that our simulation models and their configuration are close to that of the original design and that we can make reasonably accurate predictions based on these models.

\subsection{Optimization}
\begin{figure}
\begin{centering}
\includegraphics[width=\columnwidth]{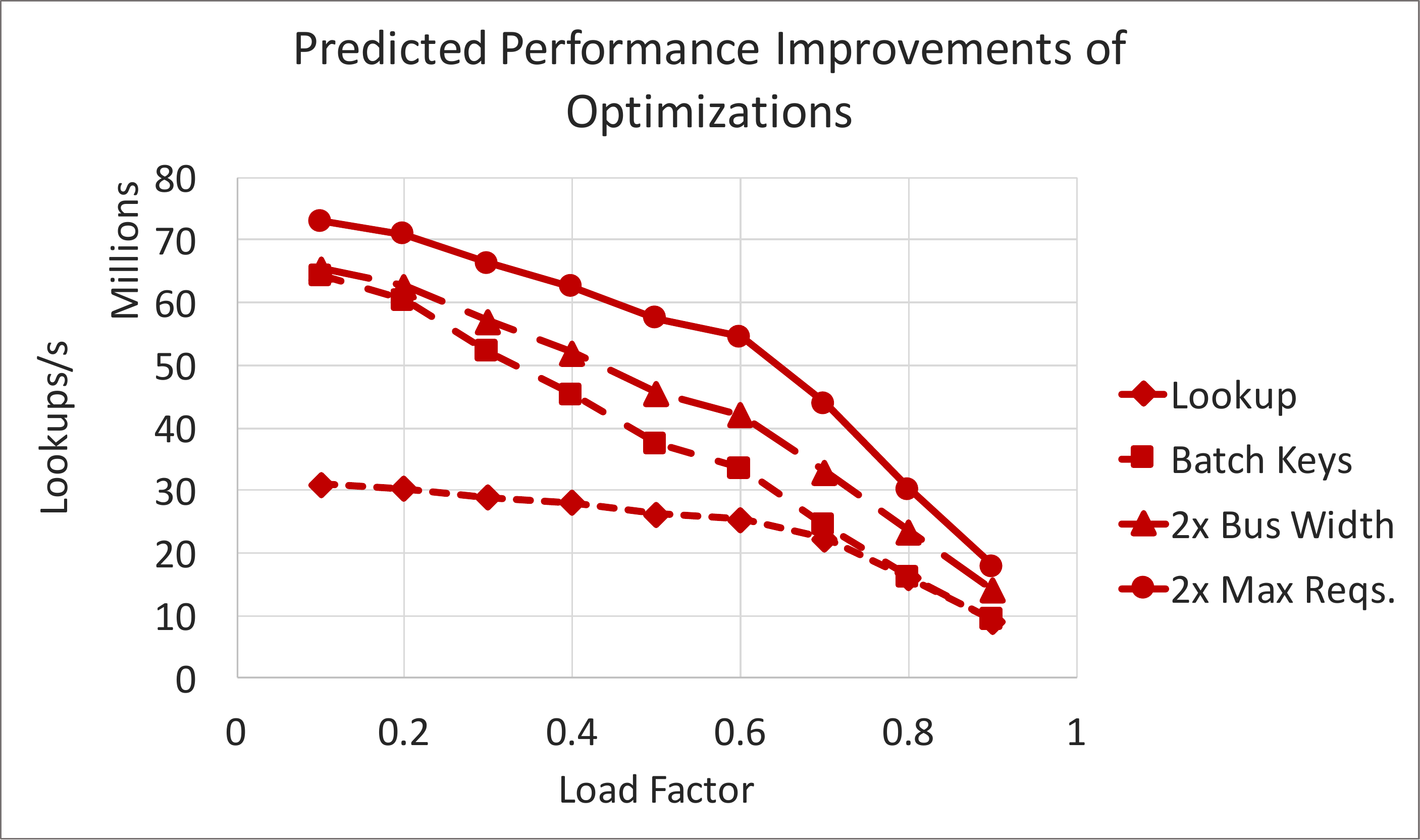}
\par\end{centering}
\caption{\label{fig:optimization} Optimization results}
\end{figure}

Once the simulation model had been validated to reproduce the original emulator hardware design accurately, we wanted to measure how much of an improvement we could obtain by adding functionality not implemented in the original hardware design. To this end, we tested the cumulative effects of the optimizations detailed in Section \ref{acc-design}. We also switched from the emulator memory model to the HMC-Sim memory model with an 85\,ns latency as it models an HMC more accurately than our simple fixed latency model. The results are shown in Figure~\ref{fig:optimization}. Each optimization in the legend includes the previous optimizations. The 2x Bus Width option includes Batch Keys, and the 2x More Requests includes both of the others.

Of the three optimizations, combining the keys into 128B packets has the biggest impact on performance (up to 2x), but only for lower load factors. These improvements likely come from the large reduction in bank conflicts, which were not accounted for by the emulator's simple memory model. Doubling the interconnect bus width also has a large impact on performance (up to 53\%), but mainly for higher load factors where hash table memory accesses are the limiting factor. Finally, increasing the number of maximum requests that the accelerator can make to memory helped some (up to 33\%), but shows that there are likely other factors that are more significant bottlenecks (like CPU overhead or accelerator clock speed). Overall, these three optimizations improve performance by 93--136\% and show that an optimized ASIC implementation of the lookup accelerator should perform significantly better than the original emulated design.

\subsection{Scaling}
\begin{figure}
\begin{centering}
\includegraphics[width=\columnwidth]{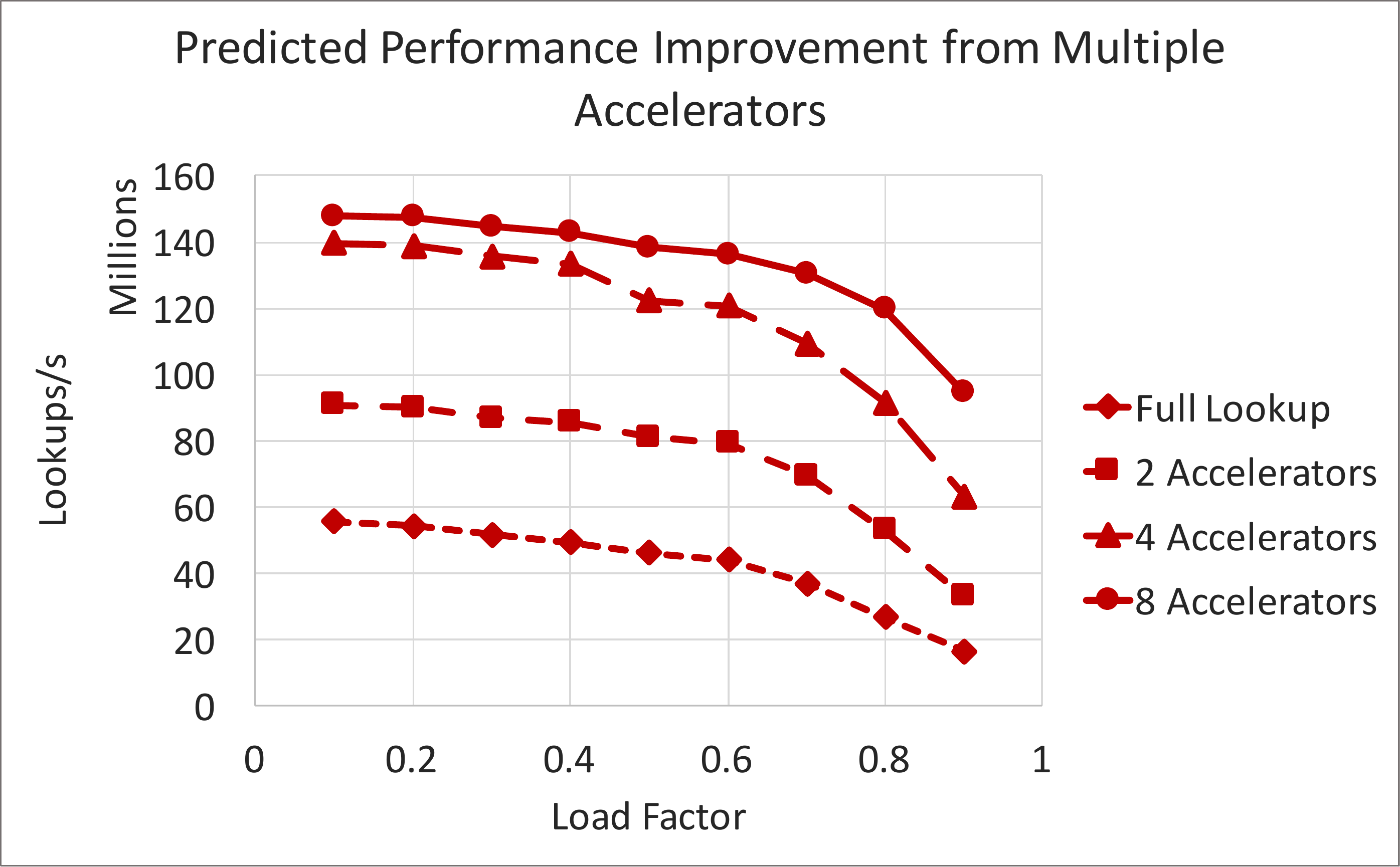}
\par\end{centering}
\caption{\label{fig:scaling} Scaling results}
\end{figure}

Even with optimizations, a single lookup accelerator cannot utilize the full bandwidth and parallelism capabilities of an HMC. For this reason, we measure how well lookup performance scales with multiple optimized lookup accelerators. For this test, we report performance in terms of full lookups to more accurately represent a normal workload that includes some work by the CPU. The results are shown in Figure~\ref{fig:scaling}.

With 2 accelerators, we see overall throughput increase by 63--100\%. Another 54--92\% can be obtained by using 4 accelerators. Finally, throughput seems to reach its limits with 8 accelerators for a total improvement of 2.7--5.7x. Interestingly, performance improvements seem to be greatest during higher load factors where more time is spent by the accelerator and CPU overhead is less significant. This suggests that memory parallelism is more of a limiting factor rather than bandwidth. To test this, we performed a second scaling experiment.

\begin{figure}
\begin{centering}
\includegraphics[width=\columnwidth]{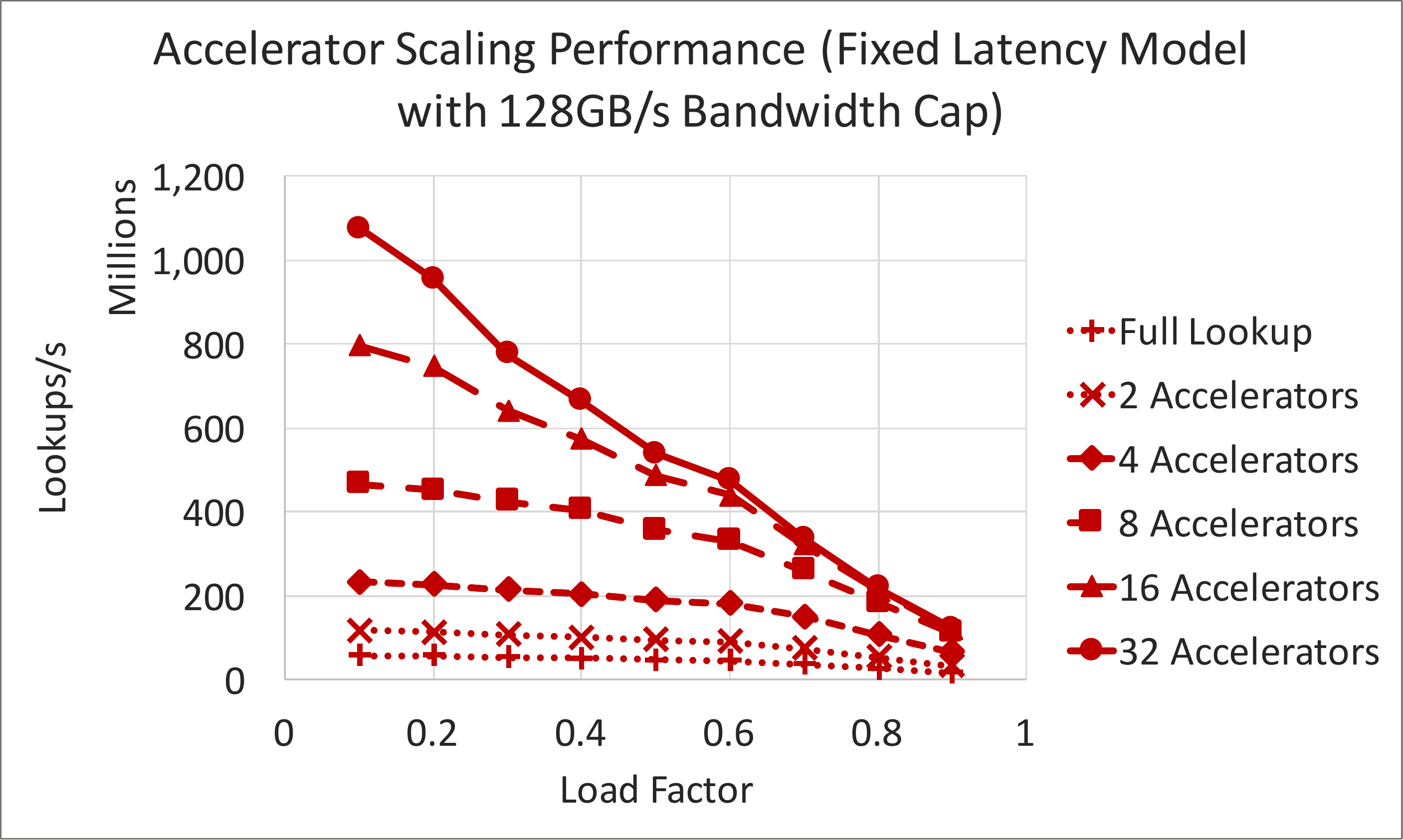}
\par\end{centering}
\caption{\label{fig:scaling2} Ideal scaling results}
\end{figure}

Since HMC-Sim simulates the vaults and banks of an HMC, it is limited by how many memory operations it can perform at any particular time. To measure scaling performance on an ideal memory with unlimited parallelism, we use our emulator memory model and increase the bandwidth limit to 128\,GB/s, which has been shown to be obtainable by existing HMCs. \cite{g112} The results are shown in Figure~\ref{fig:scaling2}.

Interestingly, single accelerator performance under the scalable emulator memory model is very close to that of the HMC-Sim memory model (0--5\% difference). When using multiple accelerators, these results show that parallelism, not bandwidth, is the limiting factor, at least for lower load factors. For higher load factors, performance is not much greater than HMC-Sim despite the increased parallelism, indicating a possible bandwidth bottleneck. However, the theoretical bandwidth of the HMC-Sim HMC model should be much greater than 128\,GB/s since it has twice the number of HMC vaults than existing HMCs and does not have the same physical constraints.

\section{Conclusions}
This work has explored the benefits of using combined hardware emulation and software simulation to explore a wider range of scenarios than possible via the emulator with better accuracy than available via simulation alone. Our results show that our simulation accurately models the original lookup accelerator design. Using a software HMC memory model enabled us to evaluate HMC-specific improvements to the query key packet stream. Experiments with improved bus width and memory parallelism were also investigated. We found that these optimizations can increase performance of a single lookup accelerator by 93--136\%. Scaling can further improve performance by another 2.7--5.7x. Overall, the simulations indicate a potential for a 5--13x speedup over the original lookup engine design and configuration.

\section{Acknowledgments}
\begin{sloppypar}   
This work was performed under the auspices of the U.S. Department of 
Energy by Lawrence Livermore National Laboratory under contract No. 
DE-AC52-07NA27344. This work was supported by Lawrence Livermore 
National Laboratory LDRD project 16-ERD-005. LLNL-CONF-738643.
\end{sloppypar}

\bibliographystyle{abbrv}
\bibliography{paper}

\end{document}